\documentclass[prd,twocolumn,reprint,preprintnumbers,nofootinbib,superscriptaddress]{revtex4-1}
\pdfoutput=1
\usepackage{cancel}

\RequirePackage[colorlinks=true
,urlcolor=blue
,anchorcolor=blue
,citecolor=blue
,filecolor=blue
,linkcolor=blue
,menucolor=blue
,linktocpage=true
,pdfproducer=medialab
,pdfa=true
]{hyperref}

\usepackage{amsmath,amssymb,amsthm,amsfonts}
\usepackage{graphicx,tabularx}
\usepackage{color}
\usepackage{multirow}
\usepackage{comment}
\usepackage{enumitem}
\usepackage{cleveref}
\usepackage{slashed}
\usepackage[normalem]{ulem}
\usepackage{scalerel}
\setlist[description]{leftmargin=0.3cm}
\setlist[itemize]{leftmargin=0.5cm}

\newcommand{\be}{\begin{equation} \begin{aligned}}
\newcommand{\ee}{\end{aligned} \end{equation}}
\newcommand{\beqa}{\begin{eqnarray}}
\newcommand{\eeqa}{\end{eqnarray}}

\def\figureautorefname~#1\null{Fig.\,#1\null}
\def\tableautorefname~#1\null{Tab.\,#1\null}

\def\equationautorefname~#1\null{Eq.\,(#1)\null}

\RequirePackage[normalem]{ulem}

\crefname{section}{Sec.}{Secs.}
\crefname{figure}{Fig.}{Figs.}
\crefname{equation}{Eq.}{Eqs.}
\crefname{appendix}{Appendix}{Appendices}

\definecolor{kjkblue}{rgb}{0.39, 0.589, 0.6914}

\def\Fermilab{Theoretical Physics Department, Fermilab, P.O. Box 500, Batavia, IL 60510, USA}
\def\CERN{Theoretical Physics Department, CERN, Esplande des Particules, 1211 Geneva 23, Switzerland}
\def\TAMU{Department of Physics and Astronomy, Mitchell Institute for Fundamental Physics and Astronomy, Texas A\&M University, College Station, TX 77843, USA}

\begin{document}

\preprint{FERMILAB-PUB-23-143-T, CERN-TH-2023-056, MI-HEE-800}

\title{
There and back again:\\
Solar cycle effects in future measurements of low-energy atmospheric neutrinos 
}

\author{Kevin J. Kelly}
\email{kjkelly@tamu.edu}
\affiliation{\CERN}
\affiliation{\TAMU}

\author{Pedro A.~N. Machado}
\email{pmachado@fnal.gov}
\affiliation{\Fermilab}

\author{Nityasa Mishra}
\email{nityasa\_mishra@tamu.edu}
\affiliation{\TAMU}

\author{Louis E. Strigari}
\email{strigari@tamu.edu}
\affiliation{\TAMU}

\author{Yi Zhuang}
\email{yiz5@tamu.edu}
\affiliation{\TAMU}

\date{\today}

\begin{abstract}
We study the impact of time-dependent solar cycles in the atmospheric neutrino rate at DUNE and Hyper-Kamiokande (HK), focusing in particular on the flux below 1~GeV.
Including the effect of neutrino oscillations for the upward-going component that travels through the Earth, we find that across the solar cycle the amplitude of time variation is about  $\pm5\%$ at DUNE, and $\pm 1\%$ at HK.
At DUNE, the ratio of up/down-going events ranges from $0.45$ to $0.85$, while at HK, it ranges from $0.75$ to $1.5$. 
Over the 11-year solar cycle, we find that the estimated statistical significance for observing time modulation of atmospheric neutrinos is $4.8\sigma$ for DUNE and $2.0\sigma$ for HK. 
Flux measurements at both DUNE and HK will be important for understanding systematics in the low-energy atmospheric flux as well as for understanding the effect of oscillations in low-energy atmospheric neutrinos. 
\end{abstract}

\maketitle
\flushbottom

\section{Introduction}\label{sec:Introduction}


\par While one of the oldest sources of neutrinos studied in high energy physics, atmospheric neutrinos continue to be an active area of research, significantly contributing to the determination of oscillation parameters~\cite{Super-Kamiokande:1998kpq, Super-Kamiokande:2010orq, Super-Kamiokande:2005mbp, Super-Kamiokande:2015qek, Super-Kamiokande:2017yvm, IceCube:2017lak, Super-Kamiokande:2019gzr}. Atmospheric neutrinos are produced when cosmic rays (CRs) collide with the Earth's atmosphere, resulting in the production of charged mesons which, through a series of decays, ultimately leading to a large flux of electron, muon, and even tau neutrinos. The neutrinos are produced ranging from sub-MeV to greater than PeV energies, and have been studied by many experiments~\cite{Gaisser:2002jj}. 

\par
\begin{figure}[t]
    \includegraphics[width = \linewidth]{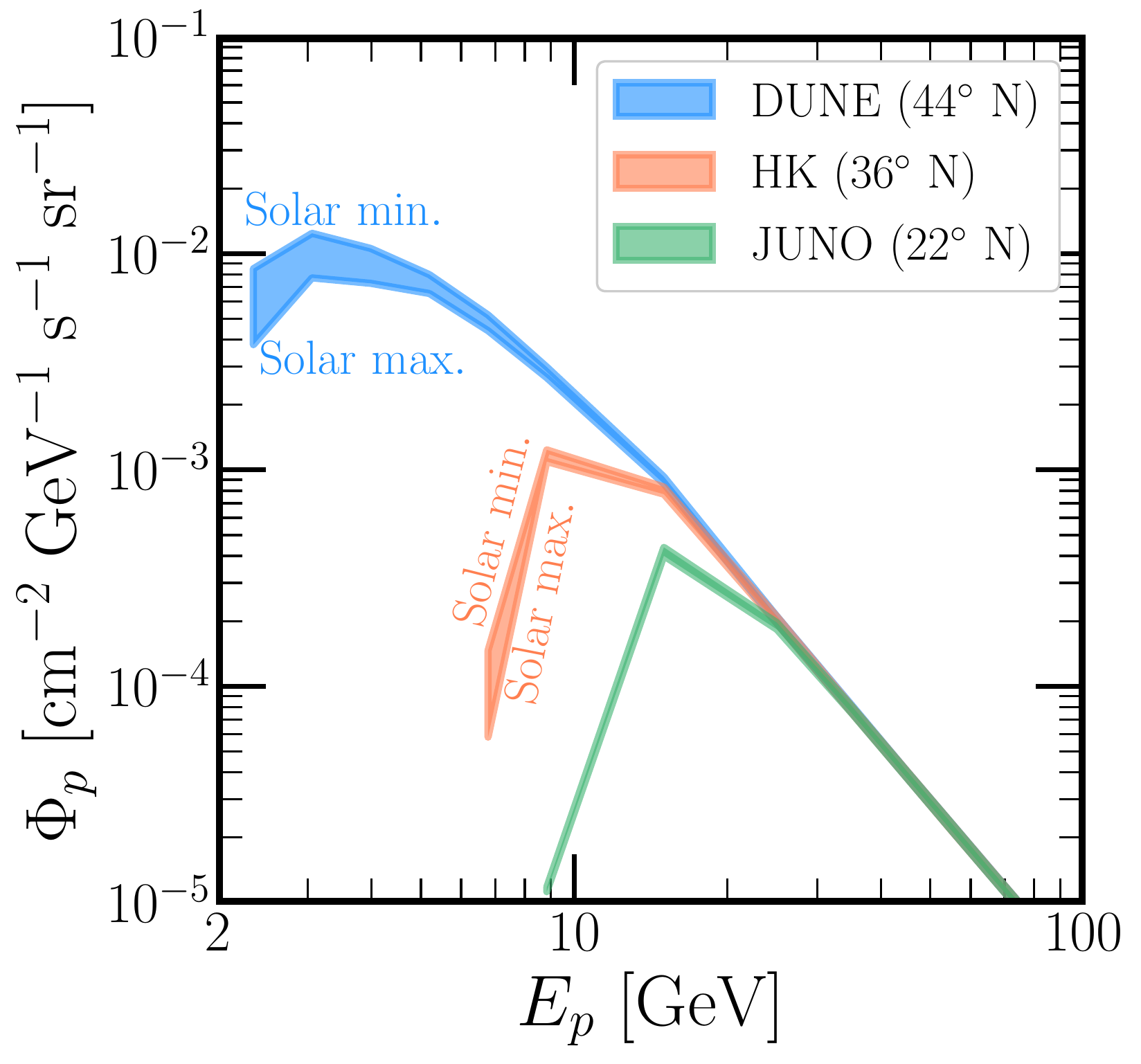}
    \caption{Cosmic ray protons producing neutrinos $E_\nu > 100$ MeV. For each spectrum (latitudes indicated), the shaded band represents the differences between solar min and solar max.  
    \label{fig:Kamioka}}
\end{figure}

 The lowest energy component of the atmospheric neutrino flux arises from CR with energies of $\lesssim 10$ GeV, see Fig.~\ref{fig:Kamioka}. 
At these energies, two distinct physical mechanisms affect the CR flux at Earth. 
First, CRs diffuse through the solar wind~\citep{Gleeson:1968zza}, so there is an expected modulation from the solar cycle~\cite{Potgieter:2013pdj}.  
This CR modulation due to solar activity has been measured by PAMELA~\cite{Marcelli:2020uqv} and BESS~\cite{Abe:2015mga}. 
The Super-Kamiokande (SK) experiment searched for such a correlation in the atmospheric neutrino flux over 20-year-long periods, but  reported a detection with a statistical significance of only 1.1$\sigma$~\cite{Super-Kamiokande:2015qek}. 
One of the challenges in measuring this effect is that it is largest at low energies, particularly below the GeV scale, and it depends on the incoming direction of the neutrinos. 
Reconstructing energy and direction of sub-GeV atmospheric neutrinos is a challenge for large Cherenkov detectors since low energy protons do not emit Cherenkov light.

\par A second mechanism that affects the low-energy atmospheric neutrino flux is the rigidity (momentum/charge) cutoff which results from the geomagnetic field. 
This rigidity cutoff is different for each location on Earth, so that the low-energy CR spectrum is different for different locations on Earth. This cut-off ultimately induces an asymmetry in the low-energy atmospheric flux~\cite{Lipari:2000du}. This effect has also been studied by SK~\cite{Super-Kamiokande:2015qek}, which identified an east-west asymmetry due to the geomagnetic field. 

The next generation of neutrino experiments will provide improved sensitivity to low-energy atmospheric neutrinos, allowing for better estimates of the solar modulation and geomagnetic field effects. 
The Deep Underground Neutrino Experiment (DUNE)~\cite{DUNE:2020ypp}, which uses liquid argon time projection chamber technology, is expected to measure sub-GeV atmospheric neutrinos and provide unparalleled reconstruction of both energy and direction of this sample~\cite{Kelly:2021jfs}. 
The Hyper-Kamiokande (HK) experiment~\cite{Hyper-Kamiokande:2018ofw}, a massive 200~kton water detector that is ten times larger than SK, is expected to have a large statistical sample that is sensitive to even percent-level time modulations of the atmospheric neutrino flux. 
The Jiangmen Underground Neutrino Observatory (JUNO)~\cite{JUNO:2021vlw}, a 20 kton liquid scintillator experiment designed to measure reactor neutrinos, is also expected to contribute to the observation of various other sources of neutrinos, including the sun, the atmosphere, and (along with DUNE and HK) the diffuse background of supernova neutrinos~\cite{Moller:2018kpn}. 
In addition to the above dedicated neutrino experiments, a future large scale dark matter detector will be somewhat sensitive to low energy atmospheric neutrinos~\cite{Aalbers:2022dzr}. 

\par In this paper, we examine the correlation between solar magnetic activity and the atmospheric neutrino flux, and the prospects for measuring this correlation at the aforementioned experiments. Despite being relatively well understood in theory, this effect has yet to be observed in neutrino experiments. 
We estimate the sensitivity of experiments to the correlation between solar magnetic activity and atmospheric neutrino flux using simulations of the atmospheric neutrino flux and detector responses in an event-by-event basis. 
By examining how the nature of the signal changes at different locations, we show that a measurement at both DUNE and HK is necessary to best understand the systematic uncertainties in the flux.

\section{Low-Energy Atmospheric Neutrino Flux}\label{sec:Fluxes}

\par
\begin{figure}[t]
    \includegraphics[width=\linewidth]{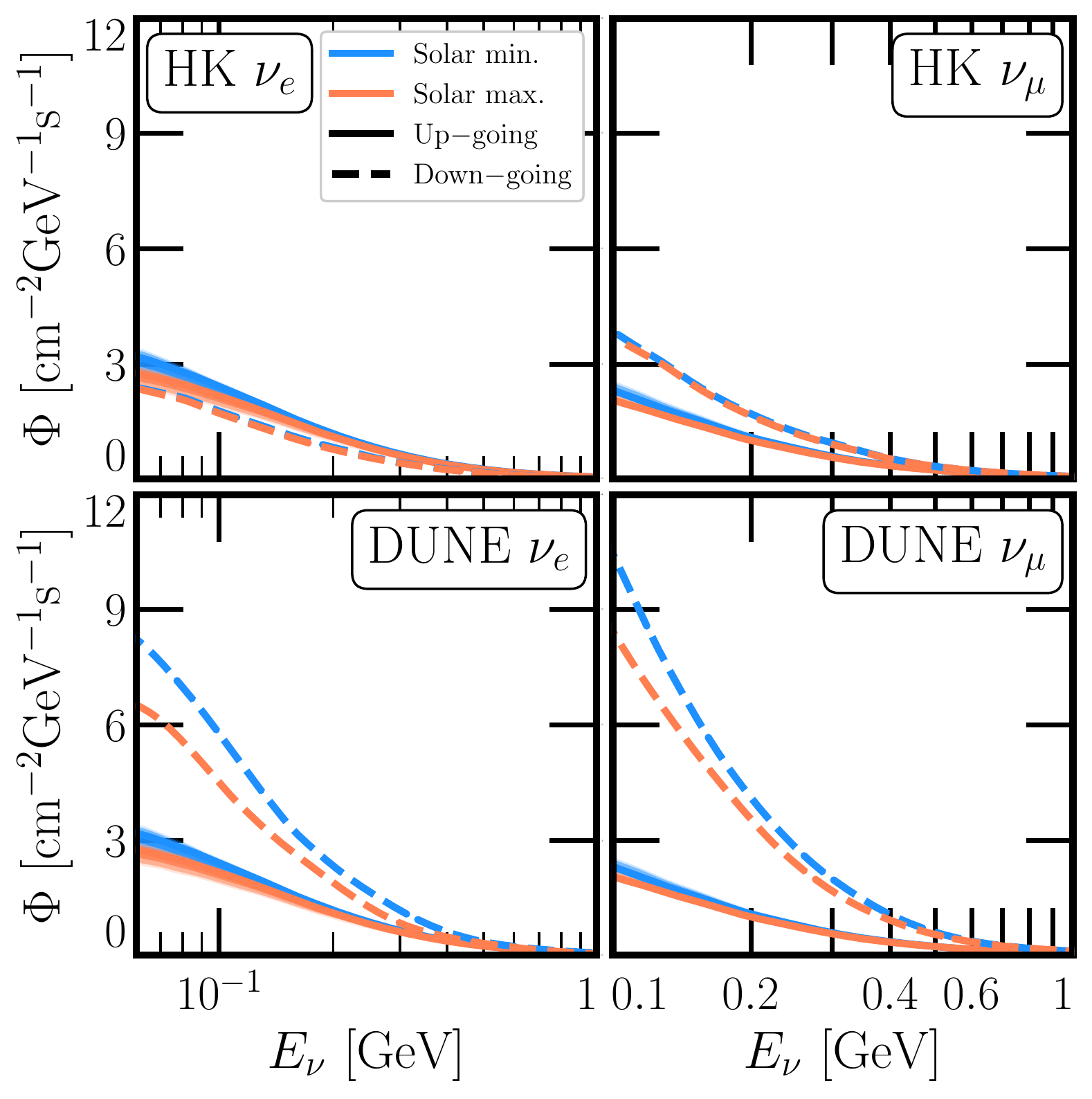}
    \caption{Electron-neutrino (left) and muon-neutrino (right) fluxes at the Hyper-Kamiokande (top) and DUNE (bottom) detectors at the extrema of the solar cycle: solar max. in blue and min. in orange. We divide the fluxes into the down-going (dashed lines) and up-going (solid bands, driven by uncertainty in neutrino oscillation parameters) components.}
    \label{fig:DUNEHKFluxes}
\end{figure}

 To calculate the low-energy atmospheric neutrino flux spectrum from the interactions of CRs, we follow the method of Ref.~\cite{Zhuang:2021rsg}. 
In Fig.~\ref{fig:DUNEHKFluxes}, we show the $\nu_e$ and $\nu_\mu$ fluxes at HK and DUNE for upgoing and downgoing neutrinos at the extrema of the solar cycles. 
In the following, we highlight relevant aspects of the calculation, and refer to  Ref.~\cite{Zhuang:2021rsg} for further details on the simulations. 
We start from the CORSIKA code~\cite{Wentz:2003bp}, which generates neutrinos from simulations of CR interactions and the subsequent air showers. 
Within CORSIKA, we use the FLUKA model to simulate low energy events, $<$ 80 GeV, and QGSJET 01C for higher energy events. 
For the input CR spectrum, we use the updated measurements of the CR spectrum on Earth at the different phases of the Solar cycle~\cite{Marcelli:2020uqv,Abe:2015mga}. 

\par We consider the detection prospects of the low-energy atmospheric neutrino flux at two future, large neutrino detectors that  will be coming online with particular strengths (and complementarity). DUNE~\cite{DUNE:2020ypp} which is at a latitude ${\sim}44^\circ$ N and has a goal of 40 kt liquid argon fiducial mass, can detect charged-current electron-neutrino scattering down to $\mathcal{O}$(tens) of MeV thanks to its liquid argon time-projection-chamber technology, as well as charged-current muon-neutrino scattering above the $\mu^\pm$ production threshold. 
Hyper-Kamiokande~\cite{Hyper-Kamiokande:2018ofw}, which is at latitude ${\sim} 36^\circ$ N, offers a significantly larger detector volume (of water), however due to Cerenkov thresholds of charged particles, its low-energy capabilities are relatively weaker than that of DUNE. 
Though our primary analysis will be presented for DUNE and HK, we also compare to the prospects for JUNO~\cite{JUNO:2021vlw}, which is at latitude ${\sim} 22^\circ$ N, and will be able to detect low-energy $\bar{\nu}_e$ events via inverse-beta-decay scattering in its liquid scintillator.

\par At any location on Earth, the neutrino flux is a sum of a downward-going component, $0 < \cos \theta_z < 1$, and an upward-going component, $-1 < \cos \theta_z < 0$, where $\theta_z$ is the zenith angle. The downgoing flux may be estimated from interactions of neutrinos in the atmosphere above the horizon. On the other hand, the upward flux requires information on the rigidity cutoff at all positions for all directions below the horizon. For this reason, estimating the upward-going flux presents a more substantial computational challenge. 
We therefore explicitly divide our simulations up into the downward and the upward flux components, and estimate the integrated neutrino fluxes over all angles for both directions. 

\par To simulate the upward-going flux for each location, we divide the Earth into 20 zenith and 20 azimuthal patches. These upward-going samples are divided by different $\cos\theta_z$ and energy so that the effects of neutrino oscillations, as described below, may be properly included -- our end result is the direction-integrated upward-going neutrino flux of each relevant neutrino flavor, $\nu_e$, $\bar\nu_e$, $\nu_\mu$, and $\bar\nu_\mu$. To check our differential fluxes as a function of zenith angle, for a fixed zenith angle we integrate over azumith, and ensure that the flux smoothly matches the fluxes from previous calculations at a zenith angle of $\cos \theta_z = 0.5$~\cite{Honda:2015fha} for neutrino energies $> 1$ GeV.

\par Upward-going neutrinos additionally experience matter-induced oscillations as they travel through the Earth. 
We calculate the oscillation probabilities for $-1 \le \cos \theta_z \le 1$ and $100$ MeV $\le E_\nu \le 1$ GeV, assuming the PREM Earth Density Model~\cite{Dziewonski:1981xy}. 
Previously, it has been demonstrated that measurements of these oscillations can provide additional, complementary information on leptonic CP-violation, specifically at DUNE~\cite{Kelly:2019itm}. 
We additionally allow the six oscillation parameters to vary assuming current knowledge of their values\footnote{Specifically, $\sin^2\theta_{12}$, $\sin^2\theta_{13}$, and $\Delta m_{21}^2$ are drawn from their assumed-Gaussian distributions, and $\sin^2\theta_{23}$, $\Delta m_{31}^2$, and $\delta_{\rm CP}$ are drawn with weights according to the $\chi^2$ table (including Super-Kamiokande atmospheric data) provided from Ref.~\cite{nufit} to include proper parameter correlations.} from Ref.~\cite{Esteban:2020cvm}. 
Note that we focus on the current knowledge of oscillation parameters because, while DUNE and HK will greatly improve these measurements with beam neutrinos and JUNO with reactor ones, it will still be invaluable to perform independent measurements with atmospheric neutrinos.
We find that varying $\sin^2\theta_{23}$ and $\delta_{\rm CP}$ according to present-day uncertainty leads to the greatest variance in expected upward-going fluxes.

\par The time variation of the flux is manifest when examining the spectrum of CR protons that produce low-energy, $> 100$ MeV neutrinos at each detector location. 
These spectra are shown in Figure~\ref{fig:Kamioka}. 
As is apparent, DUNE is more sensitive to lower-energy CRs than any other detector, sampling CRs down to the limit of our calculation of ${\sim}2$ GeV CR kinetic energy. 
For comparison, the CR spectra at HK and at JUNO cut off at higher energies, ${\sim}6$ and ${\sim}9$ GeV, respectively. 
Because of this lower energy cut-off, there is a larger variation in the CR proton flux in one solar cycle at DUNE. 

\par The corresponding low-energy neutrino fluxes are shown in Figure~\ref{fig:DUNEHKFluxes}, for DUNE and HK. Analogous fluxes for antineutrinos are presented in the appendix. 
The first interesting point to note from Figure~\ref{fig:DUNEHKFluxes} is that the upward-going fluxes at HK and DUNE are very similar. 
This is because the upgoing flux is integrated over a wide range of angular directions, so that the rigidity information in the CR spectrum is in essence averaged out. 
By similar reasoning, the impact of oscillations is similar for the DUNE and HK fluxes (the thickness of the colored bands indicate $\pm1/2/3\sigma$ uncertainty of the upward-going fluxes due to oscillation parameter uncertainty). 
The situation is different, however, for the downgoing fluxes. 
In comparison to the results from DUNE, the fluxes at HK are lower and have less variation when comparing the Solar cycle maximum and minimum.

\par A second interesting feature to note from Figure~\ref{fig:DUNEHKFluxes} is the relationship between the upward-to-downward going fluxes at each location. 
Given the nature of the magnetic field nearer to the poles, the rigidity cut-off for CRs at DUNE's Homestake Mine location is lower than it is for detectors closer to the equator such as Kamioka. 
This implies that the downgoing neutrino flux at Homestake at low energy is larger than it is for HK, which results from the CR spectrum show in Figure~\ref{fig:Kamioka}. 
However, for the upward going flux, the situation changes. 
Since Kamioka is on the opposite hemisphere relative to the South Atlantic Anomaly, which is a region defined by a low rigidity cutoff, at this location the upward going flux is larger than the downgoing flux. On the other hand, at the Homestake location, the downward going flux is larger, since on average the upward flux arises from regions of higher rigidity. 
Since the upward/downward-going ratio is different for each location, a flux measurement at both locations is crucial for understanding the systematics that arise due to the solar modulation and the geomagnetic field. 

\begin{figure}[t]
    \includegraphics[width = 0.48\textwidth]{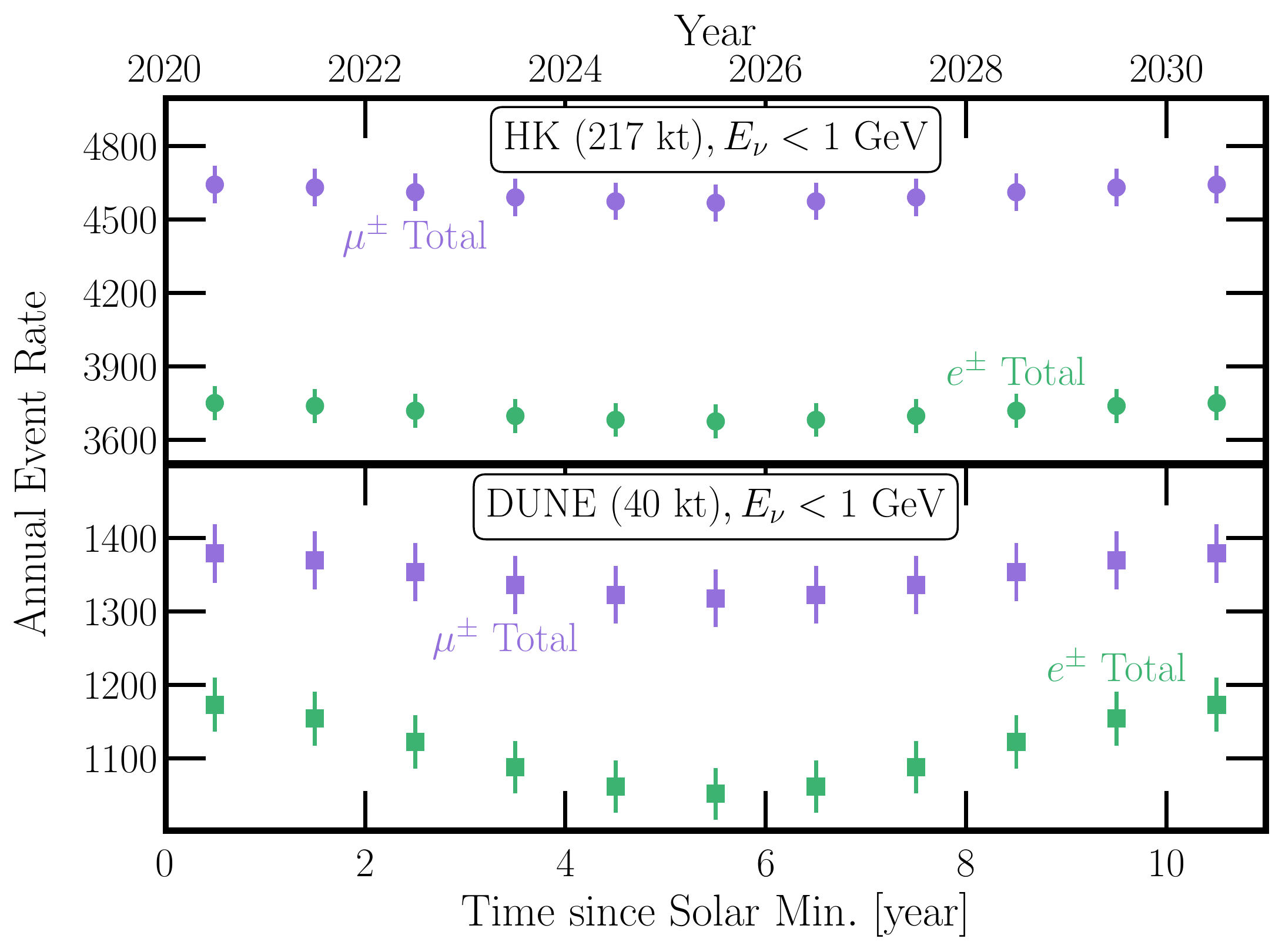}
    \caption{Event rates per year at HK (top) and  DUNE (bottom) for $\mu^\pm$ (purple) and $e^\pm$ (green) event signatures. Up-going and down-going event rates are added for simplicity.
    \label{fig:EventRates}}
\end{figure}
\section{Measurement Capabilities}\label{sec:Measurements}
\begin{figure*}[!htbp]
    \includegraphics[width = 0.96\linewidth]{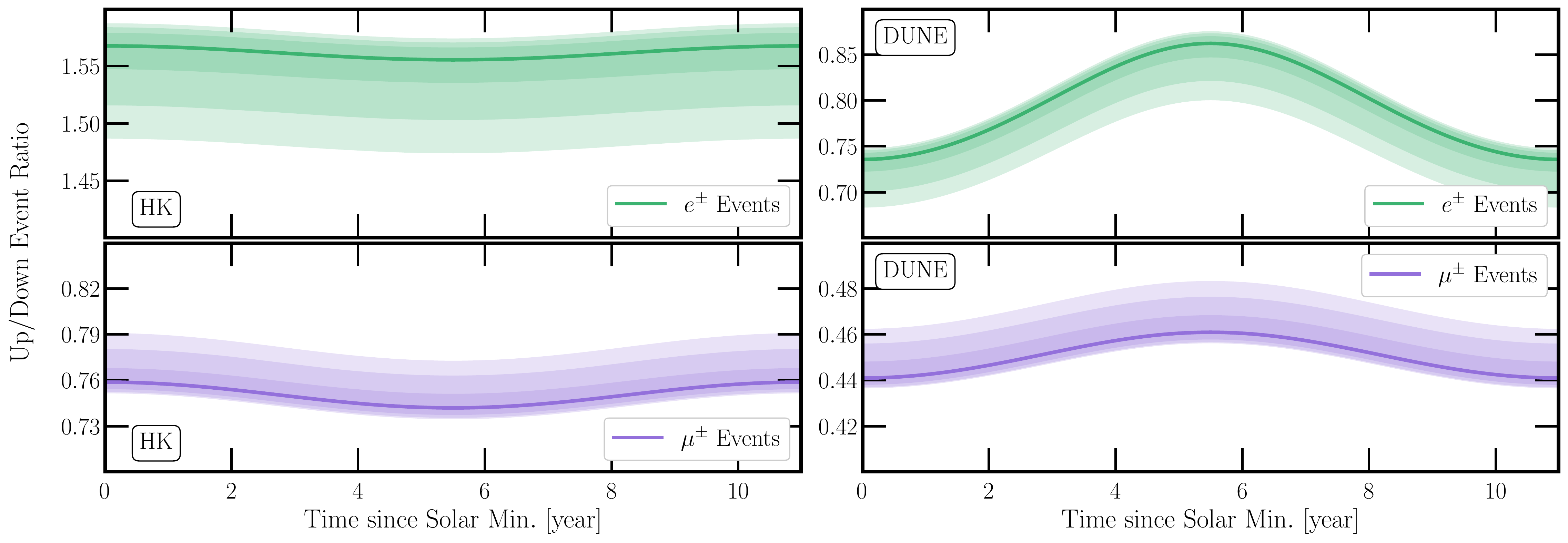}
    \caption{Ratio of up-going events to down-going ones in HK (left) and DUNE (right) over the course of one solar cycle -- the top (bottom) panels demonstrate this ratio for $e^\pm$ ($\mu^\pm$) events. In each panel, the dark/medium/light shaded regions are the $\pm 1/2/3\sigma$ allowed regions of this prediction when we vary oscillation parameters consistent with current measurements~\cite{Esteban:2020cvm}, and the thick, solid line is the median expectation. The predicted solar-cycle dependence is highly correlated once a given set of oscillation parameters is assumed to be true.
    \label{fig:DUNEUpDown}}
\end{figure*}

\par We now move on to estimating the prospects for measurement of the time variation of the atmospheric neutrino flux at DUNE and HK.\footnote{We have also determined these charged-current scattering rates for low-energy neutrinos at JUNO and find that they are too small to be meaningfully measured to the same level as at DUNE and HK, consistent with the findings of Ref.~\cite{JUNO:2021vlw}.} 
We use the \textsc{NuWro}~\cite{Golan:2012rfa,Golan:2012wx} Monte Carlo event generator to determine charged-current-inclusive scattering cross sections of the different neutrino flavors on the different detector targets at these low energies and in turn, the expected event rate in a given time period. 
For simplicity, we first consider the total $e^\pm$ and $\mu^\pm$ event rates (assuming no particle mis-identification and no charge identification) during one complete solar cycle at DUNE with a 40 kt fiducial detector volume and HK with a 217 kt one. 
The expected event rates for $E_\nu < 1$ GeV are displayed in Fig.~\ref{fig:EventRates} for $e^\pm$ (green) and $\mu^\pm$ (purple), where we take the median expected event rate subject to neutrino oscillation uncertainties and provide statistical error bars. 
The fact that these rates are not constant with respect to time is readily apparent -- for a 40 kt detector, DUNE can, in principle, detect this modulation in the $e^\pm$ and $\mu^\pm$ samples at the ${\sim}4.8\sigma$ in one cycle. If DUNE only operates with two modules and a fiducial volume of 20 kt, this reduces to ${\sim}3.4\sigma$. 
HK, in contrast, will only have sensitivity to this modulation at the ${\sim}2.0\sigma$ level due to the smaller fractional modulation of its event rates, apparent in the top half of Fig.~\ref{fig:EventRates}. 
Further challenges exist, especially for HK, in measuring and reconstructing low-energy $e^\pm$ events due to Cherenkov threholds.
Realistic reconstruction of low-energy events will reduce HK's sensitivity to the solar cycle.


In comparing the DUNE results presented here with those of Hyper-Kamiokande, we see that the expected modulation at DUNE is relatively large (a ${\sim}$10\% effect), roughly twice that predicted in HK. However, the larger detection volume of HK leads to event rates ${\sim}$3 times as large as those in DUNE. Put together, this means that such modulations should be taken into account carefully for both experiments when measuring atmospheric neutrinos, especially as it pertains to oscillation analyses.

There is an additional important issue when performing any analysis of atmospheric neutrino oscillations with these next-generation detectors that we highlight: implementing time-dependent fluxes for both upward- and downward-going neutrino components appropriately. Typically, the data are analyzed integrated over the entire lifetime of the experiment; if this is done without consideration of the flux modulation, a biased measurement of oscillation parameters (notably $\sin^2\theta_{23}$ and/or $\delta_{\rm CP}$) may be extracted from an analysis. Fig.~\ref{fig:DUNEUpDown} demonstrates this with respect to the ratio of neutrino-scattering events from the up-going flux to the down-going flux for $e^\pm$ (top) and $\mu^\pm$ (bottom). The thick line in each panel presents the median-expected result when considering the possible values of neutrino oscillation parameters given current data~\cite{Esteban:2020cvm}, whereas the shaded regions demonstrate the $\pm 1/2/3\sigma$ range allowed. Especially focusing on the DUNE panels (right), we see that the up/down ratio will vary at the several-percent level over one solar cycle. However, a few-percent change in the up/down event ratio is similarly produced by varying $\sin^2\theta_{23}$ or $\delta_{\rm CP}$ within their current uncertainties. This difference could result in an extracted atmospheric measurement at DUNE that is inconsistent with that extracted from DUNE's long-baseline $\nu_e$ appearance measurements, if these effects are not treated carefully. We note that this is a challenging measurement to perform, given the required direction reconstruction to separate up-going and down-going events~\cite{Kelly:2021jfs}, and leave dedicated studies of this and other experimental features to future work.

It is also important to note that the modulating flux ratios between HK and DUNE shown in Fig.~\ref{fig:DUNEUpDown} are out of phase -- this is because DUNE sees modulations most significantly in its down-going neutrino fluxes (see Fig.~\ref{fig:DUNEHKFluxes}), whereas HK has nearly constant down-going neutrino fluxes and small variance in the up-going ones. A combined analysis of DUNE and HK that reveals this out-of-phase dependence would provide further support for understanding of the CR flux, the varying rigidity cutoffs, and the modulation due to the solar cycle.

\section{Discussion \& Conclusions}\label{sec:Conclusions}
\par We have demonstrated that the time-variation of low-energy atmospheric neutrinos at next-generation experiments is a significant effect. This must be carefully accounted for to yield accurate predictions for neutrino-oscillation studies, and for them to be competitive with alternate methods of measuring parameters, such as the CP-violating phase in the lepton sector. This effect can be measured at DUNE and, to a lesser extent, HK, over the course of 11 years of data collection. The event rates at JUNO will be too small for meaningful measurements of these effects.

\par Interestingly, the nature of the modulation is unique at both detectors. 
DUNE sees a large modulation in its downward-going flux due to the low rigidity cutoff at its high latitude, whereas the downward-going flux at HK is fairly stable at a lower detector latitude. 
Both detectors, when viewing upward-going neutrinos, effectively sample the same profile of latitudes and longitudes. 
This, combined with the impact of neutrino oscillations in the upward-going fluxes, results in fairly similar upward-going modulation for both detector locations. 
These effects together can predict non-trivial event modulations for both electron-like and muon-like events, for both upward- and downward-going neutrinos. 
The ratio of upward-to-downward-going events is a key indicator of these effects, and we have highlighted how the out-of-phase variation of this quantity, as measured by DUNE and Hyper-Kamiokande, is a clear indicator of the solar cycle modulation. 
Further, this up-to-down-going ratio is important for the extraction of neutrino oscillation parameters and so care is required when performing these upcoming analyses.

\par We have presented results focused on generator-level information of incoming neutrinos at these two detector locations. 
On one hand, reconstructing the incoming direction of incoming sub-GeV atmospheric neutrinos is a non-trivial challenge.
While LArTPCs may be able to achieve 20-30$^\circ$ directional reconstruction~\cite{Kelly:2019itm,Kelly:2021jfs}, water Cherenkov detectors have much worse directional capabilities due to the fact that low energy protons do not emit Cherenkov light.
Determining the feasibility of realistic measurements of these modulations and up-to-down ratios, as well as more in-depth statistical measures of these time-variations, is therefore left for future work.

\par Nevertheless, measurements of neutrino oscillations will improve between now and these experiments' data collections. 
These measurements will further inform our understanding of neutrino oscillations, and we have demonstrated the importance of understanding these modulations when using atmospheric neutrinos for oscillation measurements. 
Combined measurements of DUNE and Hyper-Kamiokande should meaningfully identify such solar-cycle effects and help for a better understanding of these low-energy cosmic ray fluxes as well.



\begin{acknowledgments}
L.S., N.M., and Y.Z. are supported by the DOE Grant No. DE-SC0010813. Fermilab is managed by the Fermi Research Alliance, LLC (FRA), acting under Contract No. DE-AC02-07CH11359.

\end{acknowledgments}




\bibliographystyle{JHEP}
\bibliography{refs}

\appendix
\section{Upward- and downward-going antineutrino fluxes}
\begin{figure}[!htbp]
    \includegraphics[width=\linewidth]{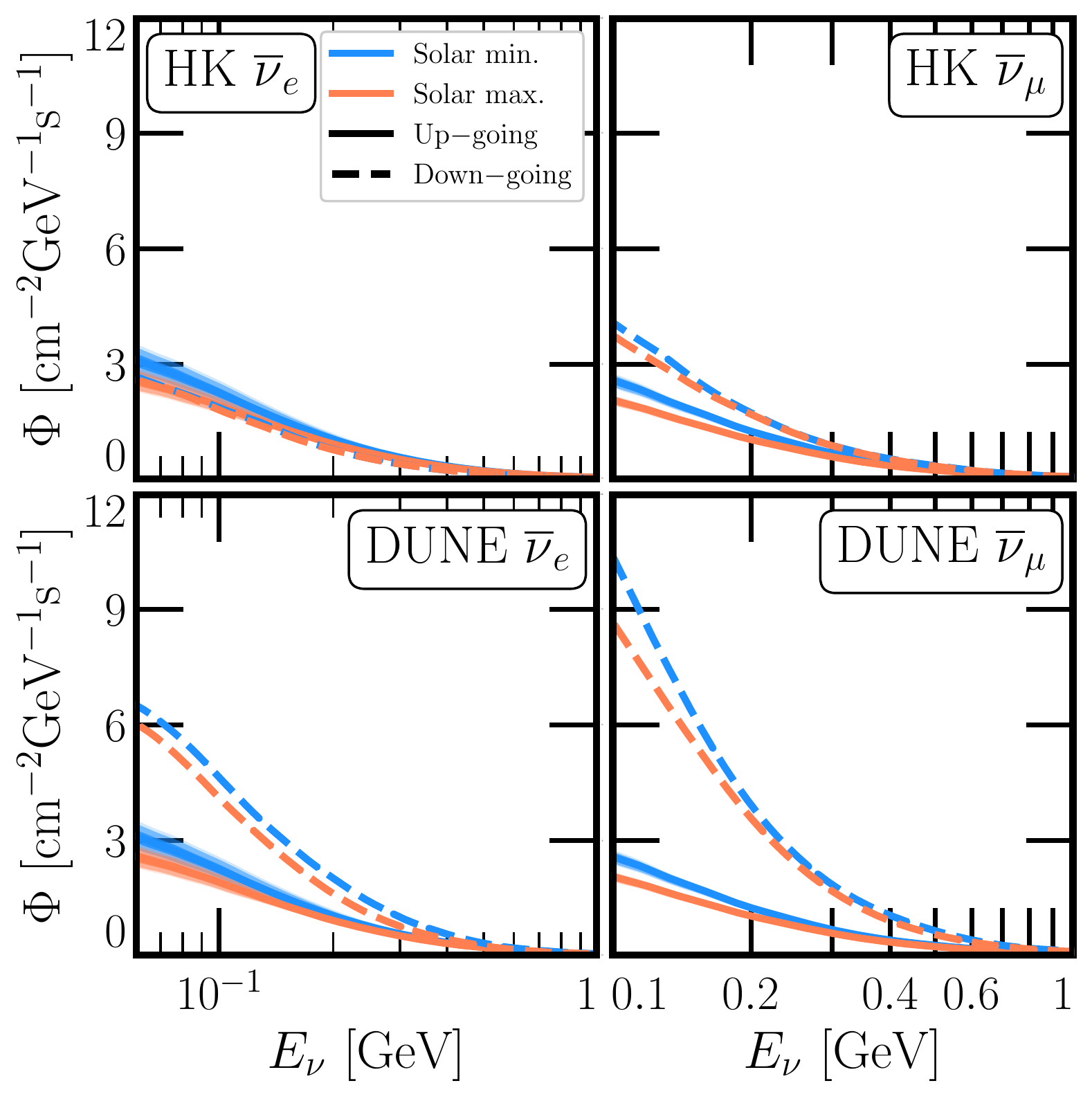}
    \caption{Electron-antineutrino (left) and muon-antineutrino (right) fluxes at the Hyper-Kamiokande (top) and DUNE (bottom) detectors at the extrema of the solar cycle: solar max. in blue and min. in orange. We divide the fluxes into the down-going (dashed lines) and up-going (solid bands, driven by uncertainty in neutrino oscillation parameters) components. See text for more detail.}
    \label{fig:DUNEHKFluxesAnti}
\end{figure}

For completeness, in Fig.~\ref{fig:DUNEHKFluxesAnti}, we show the $\overline{\nu}_e$ and $\overline{\nu}_\mu$ fluxes at HK and DUNE for upgoing and downgoing neutrinos at the extrema of the solar cycles.

\end{document}